\documentclass[11pt,twoside]{article}
\usepackage{asp2014}

\aspSuppressVolSlug
\resetcounters

\begin{document}

\title{Real-time triggering capabilities for Fast Radio Bursts at the MeerKAT telescope}

\author{F.~Jankowski,$^1$
M.~Berezina,$^{2,3}$
B.~W.~Stappers,$^1$
E.~D.~Barr,$^3$
M.~C.~Bezuidenhout,$^1$
M.~Caleb,$^1$
L.~N.~Driessen,$^1$
M.~Malenta,$^1$
V.~Morello,$^1$
K.~M.~Rajwade,$^1$
S.~Sanidas,$^1$
and M.~P.~Surnis$^1$}
\affil{$^1$Jodrell Bank Centre for Astrophysics, Department of Physics and Astronomy, The University of Manchester, Manchester M13 9PL, UK; \email{fabian.jankowski@manchester.ac.uk}}
\affil{$^2$Landessternwarte, Universit\"{a}t Heidelberg, K\"{o}nigstuhl 12, D-69117 Heidelberg, Germany}
\affil{$^3$Max-Planck-Institut f\"{u}r Radioastronomie, Auf dem H\"{u}gel 69, D-53121 Bonn, Germany}

\paperauthor{Fabian~Jankowski}{fabian.jankowski@manchester.ac.uk}{0000-0002-6658-2811}{Jodrell Bank Centre for Astrophysics}{Department of Physics and Astronomy, The University of Manchester}{Manchester}{Greater Manchester}{M13 9PL}{UK}
\paperauthor{Marina~Berezina}{mberezina@lsw.uni-heidelberg.de}{}{}{Landessternwarte, Universit\"{a}t Heidelberg}{Heidelberg}{}{69117}{Germany}
\paperauthor{Benjamin~Stappers}{}{}{Jodrell Bank Centre for Astrophysics}{Department of Physics and Astronomy, The University of Manchester}{Manchester}{Greater Manchester}{M13 9PL}{UK}
\paperauthor{Ewan~Barr}{}{}{Max-Planck-Institut f\"{u}r Radioastronomie}{Auf dem H\"{u}gel 69}{Bonn}{Nordrhein-Westfalen}{53121}{Germany}
\paperauthor{Mechiel~Bezuidenhout}{}{}{Jodrell Bank Centre for Astrophysics}{Department of Physics and Astronomy, The University of Manchester}{Manchester}{Greater Manchester}{M13 9PL}{UK}
\paperauthor{Manisha~Caleb}{}{}{Jodrell Bank Centre for Astrophysics}{Department of Physics and Astronomy, The University of Manchester}{Manchester}{Greater Manchester}{M13 9PL}{UK}
\paperauthor{Laura~Driessen}{}{}{Jodrell Bank Centre for Astrophysics}{Department of Physics and Astronomy, The University of Manchester}{Manchester}{Greater Manchester}{M13 9PL}{UK}
\paperauthor{Mateusz~Malenta}{}{}{Jodrell Bank Centre for Astrophysics}{Department of Physics and Astronomy, The University of Manchester}{Manchester}{Greater Manchester}{M13 9PL}{UK}
\paperauthor{Vincent~Morello}{}{}{Jodrell Bank Centre for Astrophysics}{Department of Physics and Astronomy, The University of Manchester}{Manchester}{Greater Manchester}{M13 9PL}{UK}
\paperauthor{Kaustubh~Rajwade}{}{}{Jodrell Bank Centre for Astrophysics}{Department of Physics and Astronomy, The University of Manchester}{Manchester}{Greater Manchester}{M13 9PL}{UK}
\paperauthor{Sotiris~Sanidas}{}{}{Jodrell Bank Centre for Astrophysics}{Department of Physics and Astronomy, The University of Manchester}{Manchester}{Greater Manchester}{M13 9PL}{UK}
\paperauthor{Mayuresh~Surnis}{}{}{Jodrell Bank Centre for Astrophysics}{Department of Physics and Astronomy, The University of Manchester}{Manchester}{Greater Manchester}{M13 9PL}{UK}





\begin{abstract}
Fast Radio Bursts (FRBs) are bright enigmatic radio pulses of roughly millisecond duration that come from extragalactic distances. As part of the MeerTRAP project, we use the MeerKAT telescope array in South Africa to search for and localise those bursts to high precision in real-time. We aim to pinpoint FRBs to their host galaxies and, thereby, to understand how they are created. However, the transient nature of FRBs presents various challenges, e.g. in system design, raw compute power and real-time communication, where the real-time requirements are reasonably strict (a few tens of seconds). Rapid data processing is essential for us to be able to retain high-resolution data of the bursts, to localise them, and to minimise the delay for follow-up observations. We give a short overview of the data analysis pipeline, describe the challenges faced, and elaborate on our initial design and implementation of a real-time triggering infrastructure for FRBs at the MeerKAT telescope.
\end{abstract}

\section{Introduction}
\label{sec:introduction}

Fast radio bursts (FRBs;~\citealt{2007Lorimer, 2013Thornton}) are luminous short-duration (about ms) extra-galactic radio transients of yet unclear nature (for recent reviews see~\citealt{2019Petroff, 2019Cordes}). Observed to be apparently of two types (repeating and non-repeating), FRBs are the subject of cutting-edge research. Open questions concern their progenitors, their possible emission mechanism(s), and the relation between the two types of events, with a plethora of competing theories suggested\footnote{\url{https://frbtheorycat.org}} \citep{2018Platts}. A big step on the way to understanding the FRB mystery was the identification of host galaxies of some of the bursts\footnote{\url{https://frbhosts.org}}. While localisation of repeating FRBs is ``simply'' a matter of regular observations, the case is more complicated for the apparently one-off events. The latter demands radio interferometers with multi-beaming capabilities together with real-time processing and triggering.

We use the 64-element MeerKAT telescope array in South Africa to search for, localise, and understand the nature of FRBs. To do that, we designed and commissioned the MeerTRAP backend that performs radio transient searches in real-time and fully commensally with other projects running at the telescope.

\section{Why is real-time triggering crucial?}
\label{sec:realtimetriggering}

The vast majority of FRBs appear as single-peaked pulses at time resolutions of a few milliseconds, that are common in current single-pulse search efforts. Those searches are usually limited by the employed sampling times and intra-channel dispersive smearing. However, as first demonstrated by \citet{2018Farah}, and more recently by e.g.~\citet{2020Day, 2020Nimmo}, some FRBs exhibit intriguingly complex burst morphologies, frequency structure, and polarimetric properties at microsecond resolution. Having data with full time, frequency and polarisation information (voltage data) is crucial for those studies, not only because of the higher time resolution but also because dispersive propagation effects can be removed completely via coherent dedispersion. Those data, therefore, allow us to constrain the FRB emission model(s). The sheer volume prohibits the recording of voltage data for all candidates.

Only the phase information contained in complex voltage data makes it possible to produce synthesis snapshot images of the field surrounding an FRB in post-processing, thereby localising them with up to arcsecond precision. We can access the raw data from each telescope element pair quasi separately. The localisation ability is essential in our search for FRB host galaxies and to understand FRB progenitors. For instance, the Australian Square Kilometre Array Pathfinder (ASKAP) team have done that successfully (e.g.~\citealt{2019Bannister}).

Effective multi-wavelength follow-up can only realistically be performed if the sky locations of FRBs are known reliably, and ideally to better than a few arcseconds, much below the fields-of-view of follow-up telescopes. High-precision FRB localisations are therefore essential for follow-up by both ground and space-based facilities. Multi-wavelength and multi-messenger observing efforts help understand the emission properties of FRBs.

\section{Design and implementation at MeerKAT}
\label{sec:triggeringmeerkat}

\begin{figure}
    \centering
    \plotone[width=\textwidth]{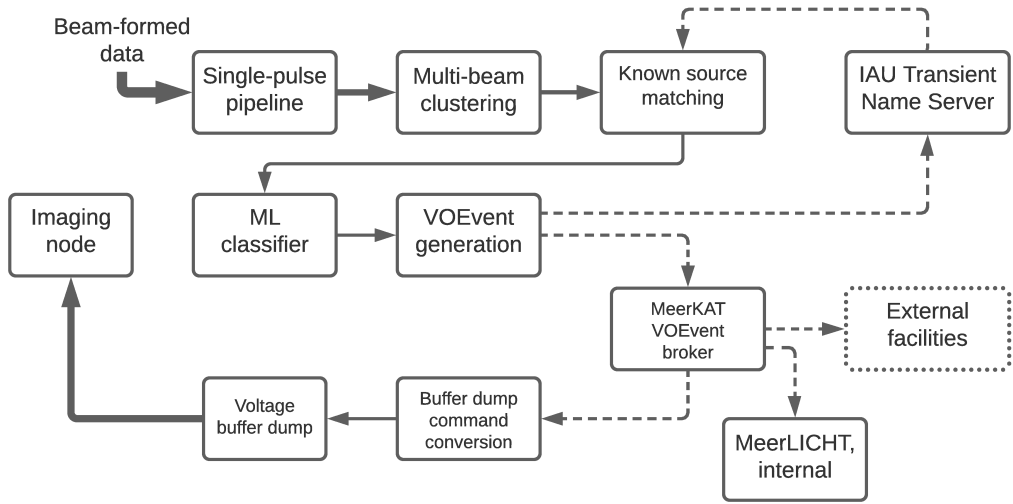}
    \caption{Abstracted overview diagram of the MeerTRAP real-time triggering pipeline at the MeerKAT telescope. The line thickness represents the approximate data rate but is not to scale.}
    \label{fig:pipeline}
\end{figure}

In Fig.~\ref{fig:pipeline} we show an overview of the MeerTRAP real-time processing infrastructure. It consists of a GPU-based single-pulse detection pipeline, for which we use the \textsc{astroaccelerate} software \citep{2019Carels}, that we wrapped and augmented. We designed and implemented a host of tools that process the candidates output by the real-time pipeline. Cornerstones to allow snapshot localisation are the transient buffers, several POSIX shared memory \textsc{psrdada}\footnote{\url{http://psrdada.sourceforge.net/}} buffers, that retain typically about 50~s of channelised complex voltage data. In the event of a burst detection in the low-resolution data, a small percentage of their content gets saved for further offline processing.

\subsection{Major challenges}
\label{sec:challenges}

The finite size of the transient buffers enforces strict real-time requirements for both our single-pulse detection pipeline and the candidate processing stack at later stages. The software must be able to process the candidate data in $< 10$ seconds after the single-pulse pipeline and in $< 40$ seconds in total. Another challenge concerns reducing the rate of single-pulse candidates caused by radio frequency interference (mainly from satellites), and known radio sources, such as pulsars. Equivalently, the problem is to find genuine astrophysical FRB events that do not originate from known radio sources, above a significant candidate noise floor.

We developed \texttt{python} tools that perform multi-beam clustering and known source matching from given catalogues,\footnote{\url{https://bitbucket.org/jankowsk/meertrapdb/}} which in combination typically reduce the candidate rate by 70 to 90~per~cent. An image-based machine-learning classifier is expected to reduce this rate further (to $< 1$~per~cent), once fully deployed.

\subsection{VOEvents}
\label{sec:voevents}

As part of the real-time triggering pipeline, we utilise VOEvent packets to initiate the write out the voltage buffers and to alert internal facilities, such as the optical MeerLICHT telescope. We chose the VOEvent format because it is a standard that is well tested and adopted by the transient community. For example, it is used to alert gamma-ray burst detections (e.g.~\textit{Swift} and \textit{Fermi} satellites), gravitational wave events by LIGO/Virgo, as well as events in other wavebands and messengers. Additionally, a VOEvent standard for FRB alerts has been developed \citep{2017Petroff} and was adopted at several radio facilities, such as the Murchison Widefield Array \citep{2019Hancock}. Finally, it will allow us to disseminate our triggers to external collaborators in the future easily.

VOEvent packets are in XML format \citep{2011Seaman} and contain the parameters of the alert, e.g.\ a unique identifier, the author, the event time, its sky position, and the instrumental setup. The event packets are distributed by brokers, for which we employ the \textsc{comet} software \citep{2014Swinbank}, both locally on the MeerTRAP head node and the MeerKAT telescope-internal broker. Our triggering software allows for easy event packet creation and emission\footnote{\url{https://bitbucket.org/jankowsk/meertrig/}}. FRB parameters, like the sky position, will be updated successively via VOEvent update alerts, as we refine them.

\section{Proof of concept and software availability}
\label{sec:proofofconcept}

We demonstrated successful early triggering using the initial system on the bright Vela pulsar, PSR~J0835$-$4510. We saved a few seconds of complex voltage data and formed snapshot synthesis images and a high time-resolution pulse profile.

Most of our software is publicly available and is linked from the MeerTRAP project website: \url{https://www.meertrap.org}.

\section{Acknowledgements}
\label{sec:acknowledgements}

MB acknowledges support from the Bundesministerium f\"{u}r Bildung und Forschung (BMBF) D-MeerKAT award 05A17VH3 (Verbundprojekt D-MeerKAT). FJ and the Manchester group acknowledge funding from the European Research Council (ERC) under the  European Union's Horizon 2020 research and innovation programme (grant agreement No. 694745).

\bibliography{P6-125}

\end{document}